\documentclass[twocolumn,10pt,showpacs,amsmath,amssymb,prl]{revtex4}

\usepackage{graphicx}
\usepackage{dcolumn}

\begin{document}

\title{Thermoelastic-damping noise from sapphire mirrors \\
         in a fundamental-noise-limited interferometer}
	
\date{\today}
\author{Eric D. Black, Akira Villar, and Kenneth G. Libbrecht}
\affiliation{LIGO Project, California Institute of Technology \\
Mail Code 264-33, Pasadena CA 91125}

\begin{abstract}

%Currently, the most advanced and sensitive interferometers in the world are limited by fundamental thermal noise in the mirrors. Schemes for beating the standard quantum limit, especially for gravitational wave detection~\cite{Buonanno01}, require new mirrors that exhibit less thermal noise. High-purity, synthetic sapphire is one candidate material for these mirrors. 

We report the first high-precision interferometer using large sapphire mirrors, and we present the first direct, broadband measurements of the fundamental thermal noise in these mirrors. Our results agree well with the thermoelastic-damping noise predictions of Braginsky, et al.~\cite{Braginsky99-2} and Cerdonio, et al.~\cite{Cerdonio01}, which have been used to predict the astrophysical reach of advanced interferometric gravitational wave detectors.

\end{abstract}


\maketitle

There is currently a large, multi-national effort to establish gravitational wave astronomy using interferometric detectors. A number of observatories are either being constructed or have recently been completed, including LIGO \cite{Barish99}, GEO \cite{Luck97}, VIRGO \cite{Caron97}, TAMA \cite{Kawabe97}, and ACIGA \cite{Blair00}. As the community builds, operates, and learns from these first-generation detectors, advanced second-generation instruments are being developed that are designed to beat the standard quantum limit and dramatically increase the astrophysical reach of these observatories~\cite{Buonanno01,LIGO2}. 

Thermal noise in the test-mass mirrors, however, sets a fundamental limit to the sensitivity of all optical interferometers. If not overcome, this noise source will prevent the standard quantum limit from even being reached, much less surpassed. Thermal noise can be reduced with the development of new mirror materials, and one especially promising candidate material for advanced mirror substrates is sapphire. 

Sapphire is expected to show lower levels of broadband thermal noise than the fused-silica mirrors of first-generation detectors~\cite{Rowan00} and to exhibit less thermal lensing~\cite{Ju96}. However, while fused-silica is relatively well-established as a mirror substrate material, to date no high-sensitivity interferometer using sapphire mirrors has been constructed. In this paper we report on the first high-sensitivity interferometer with sapphire mirrors, and we report the first observation of the fundamental noise floor set by sapphire mirror substrates.

\section{Theory}

The ultimate sensitivity of a ground-based interferometric gravitational wave detector is limited by several fundamental noise sources: quantum noise from the readout scheme (typically a combination of photon shot noise and radiation pressure noise inherent in the interferometer), thermal noise in the interferometer components, and gravity-gradient noise arising from seismic motions around the detector~\cite{Saulson-book}. These noise sources remain after a host of non-fundamental noise sources -- direct seismic noise, electronic noise, laser frequency fluctuations, etc. -- have been reduced using a variety of abatement techniques. 

%A gravitational wave detector suitable for use as an astronomical instrument needs to have extraordinary sensitivity. Its test masses need to be extremely quiet to resolve the strain induced by the weak tidal forces produced by gravitational waves of astrophysical origin. Quantum mechanics, of course, sets a fundamental limit on our ability to measure the position of any mass, be it an atom or a multi-kilogram mirror in an interferometer. However, with a clever enough measurement scheme it may be possible to circumvent this quantum limit (see, for example, Reference~\cite{Buonanno01}).
% which is an intriguing proposition by itself, whether you intend to do astronomy or not. 

%Before the quantum limit can be surpassed, or even met, in an interferometer, all other noise sources in the instrument must be made small by comparison. One fundamental noise source of special concern is thermal noise, which gives rise to fluctuations in the apparent position of a mirror that have nothing to do with quantum mechanics. %These fluctuations often occur at levels at or above the quantum noise in a gravitational wave detector, limiting our ability both to do astronomy and to develop an instrument that can circumvent the quantum measurement limit. 

Thermal noise in this application arises because an interferometer measures the position of a mirror (test mass) surface and not its true center of mass. Thermal fluctuations in the mirror substrate and optical coatings thus add noise to the interferometer readout and limit its sensitivity. These fluctuations are intimately related to losses and irreversibility in thermodynamics, as described by Onsager~\cite{Onsager31-1,Onsager31-2}, and especially the fluctuation-dissipation theorem~\cite{Callen51,Callen52}. Given a mechanical loss mechanism, these theories can be used to predict the fundamental fluctuations in the relevant thermodynamical variables. 

One such loss mechanism is thermoelastic damping, where energy is lost to heat flow in a material. As a material is flexed, regions are compressed or expanded and, in response, heat up or cool down according to Le Ch\^{a}telier's principle~\cite{LeChatelier84}. This creates a temperature gradient which drives heat flow, and dissipates some of the mechanical energy used to produce the flexing. The converse also occurs in accordance with the fluctuation-dissipation theorem. The temperature of a macroscopic object is an averaged quantity, and local fluctuations are always present, even when the object is in thermal equilibrium. These fluctuations drive mechanical noise in the material through thermal expansion. This is distinct from Brownian motion, which is noise driven by internal frictional losses~\cite{Saulson90,Numata03}.

It has been known for some time that thermoelastic damping loss can contribute to thermal noise in mechanical systems~\cite{Saulson90}. However, that this mechanism can contribute significantly to the noise floor of an interferometer with sapphire mirrors was first pointed out by Braginsky and colleagues in 1999~\cite{Braginsky99-2}. This noise source can be thought of either as thermoelastic-damping mediated thermal noise, as described by the fluctuation-dissipation theorem, or as a coupling between intrinsic temperature fluctuations in the bulk of the mirror and the mirror's thermal expansion coefficient. Either picture gives the same result, and this fundamental noise source is expected to set the ultimate sensitivity limit of an advanced gravitational wave detector that uses sapphire mirrors. 

Braginsky, et al.~\cite{Braginsky99-2} made a quantitative prediction of this noise source valid in the limit of high frequencies or large spot sizes. In this limit, heat flow is perpendicular to the surface of the mirror, and the effect can be modeled by a one-dimensional diffusion equation. This result was later extended by Cerdonio, et al.~\cite{Cerdonio01}, who solved the multi-dimensional problem and derived an expression valid at all frequencies and spot sizes. The latter gives the displacement noise power spectral density (in $m^2/Hz)$
\begin{equation}
\label{eq:Cerdonio}
S_{TE} (\omega) = \frac{8}{\sqrt{\pi}} \frac{\alpha^2 (1+\sigma)^2}{\kappa} k_B T^2 w_0 J\left( \Omega \right),
\end{equation}
where the dimensionless integral $J \left( \Omega \right)$ is given by
\begin{equation}
\label{eq:Cerdonios-integral}
J \left( \Omega \right) = \frac{\sqrt{2}}{\pi^{3/2}} \int_0^{\infty} du \int_{-\infty}^{+\infty} dv 
	\frac{u^3 e^{-u^2/2}}{(u^+v^2) [(u^2+v^2)^2+\Omega^2]}
\end{equation}
and the dimensionless frequency $\Omega$ is
\[
\Omega = \frac{\omega}{\omega_c},
\]
where
\[
\omega_c = \frac{2\kappa}{\rho C w_0^2}
\]
In these expressions, $\alpha$ is the thermal expansion coefficient of the mirror substrate, $\sigma$ is its Poisson's ratio, $\kappa$ is the thermal conductivity, $\rho$ is its mass density, $C$ is its specific heat, and $\omega$ is the (angular) measurement frequency. The laser spot radius $w_0$ we use in this paper is the usual one, where the electric field falls off to $1/e$ of its maximum value. 

The expression for $J(\Omega)$ given in the original paper~\cite{Cerdonio01} omitted a factor of $1/\pi$ in the integral $J\left(\Omega\right)$. Also Cerdonio, et al.\ and Braginsky, et al.\ use as the spot size $r_0$ the radius at which the intensity of the beam falls off to $1/e$ of its maximum value. This quantity is related to $w_0$ by $r_0 = w_0 / \sqrt{2}$.

At high frequencies $(\omega \gg \omega_c)$, this noise source has a characteristic frequency dependence $S^{1/2}(\omega) \propto 1/\omega$, and in this region Braginsky, et al.'s formula is valid. At lower frequencies, the frequency dependence is weaker than $1/\omega$. While we do not fully sample the low-frequency regime $(\omega \ll \omega_c)$ in the present experiments, we do see enough of the transition region between the two to need a theory that is valid for all frequencies.

\section{The instrument}

\begin{center}
\begin{figure}
\includegraphics{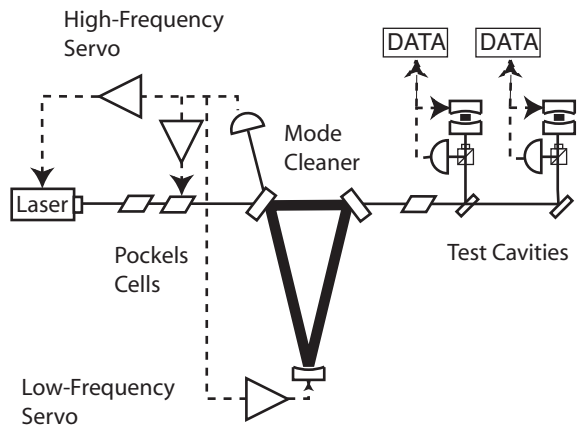}
\caption{\label{fig:layout} Schematic of the interferometer used for the present measurements. Each of the four test mass mirrors and the three mode cleaner mirrors are mounted on individual single-pendulum suspensions in a vacuum chamber. The laser is locked to the mode cleaner via a three-path servo for frequency stabilization, and the test cavities are independently locked to the stabilized beam. The test cavity lengths (1 cm each) are short, to optimize the system for measuring displacement noise. The radii of curvature of the test-mass mirrors are long (1 meter) to increase the spot size, thus reducing the displacement noise, and making the measurement more relevant to gravitational wave detection.}
\end{figure}
\end{center}

Figure~\ref{fig:layout} shows a diagram of the instrument used for these measurements. The interferometer was constructed using two short, independent arm cavities, made up of four identical mirrors. In order to reduce the thermal noise to a level relevant to gravitational wave detection, the radii of curvature of these mirrors (1 meter) were chosen to be much larger than the cavity length (1 cm), giving a relatively large spot size $(w_0 = 160\ \mu m)$ for an interferometer of this size~\cite{Black04-2}.

All four of the test-cavity mirrors and the fused-silica mode cleaner mirrors were suspended as single pendula in a vacuum chamber, with magnet-and-coil damping and actuation systems. Since cavities made from independent, suspended optics typically exhibit very low noise at high frequencies, above $12 Hz$ the mode cleaner served as the ultimate frequency reference for the experiment. There, the laser frequency followed the mode cleaner using a two-path servo, with actuation on the laser's internal PZT stabilizing the frequency noise up to $30 kHz$, and actuation on an external broadband Pockels cell acting at higher frequencies. At frequencies below $12 Hz$, the mode cleaner was locked to the laser to suppress seismic noise. In addition to a frequency reference, the mode cleaner also provided spatial filtering of the beam.

The arm cavities were locked independently to the resulting stabilized and filtered beam, with the arm-cavity servos actuating directly on the arm cavities' output mirrors and acting only at low frequencies. Data were collected from the error signals of the arm cavities, and the difference between the two data streams was calculated in real time to remove any remaining laser frequency noise or other common-mode noise. The calibration of the instrument was described in detail in a previous paper~\cite{Black04-2}.

In our previous experiments with this interferometer~\cite{Black04-2}, we measured displacement noise using fused-silica test-cavity mirrors. In this first configuration we characterized a variety of noise sources and verified that what appeared to be displacement noise really did originate inside the test cavities. For the present experiment, we replaced all four fused-silica test-cavity mirrors with sapphire mirrors of the same geometry, replacing the suspension wires at the same time with thicker ones to keep the violin mode frequencies roughly the same. Any difference, then, between the first and second noise spectra should be due only to the difference between fused-silica and sapphire substrates. The coatings on both sets of mirrors were identical, both being $\text{SiO}_2/\text{Ta}_2\text{O}_5$ coatings of the same thickness, made by the same manufacturer~\cite{REO}. Each test cavity had a finesse of approximately $10,000$ and a transmission of $70\%$.

One aspect of using sapphire mirrors became clear when we performed this measurement: Mechanical resonances in the mirrors were much less problematic than with fused-silica mirrors. With fused-silica mirrors in our interferometer, we needed eight notch filters in the arm-cavity servos to keep the lowest-frequency mirror resonances from ringing up. The same resonances in sapphire mirrors occur at nearly twice the frequencies of those in fused-silica mirrors. Moreover, even though the intrinsic $Q$'s of sapphire mirrors are expected to be higher than those of fused-silica optics, the \emph{in-situ} $Q$'s are largely determined by the suspensions and are not much higher than those of the fused-silica optics. These two effects combined to remove the mirror modes far enough away, in frequency, from the unity-gain frequency of the servo that no notch filters were required. The interferometer thus operated much more stably with sapphire mirrors than with fused-silica.

\section{Results}

\begin{center}
\begin{figure}
\includegraphics{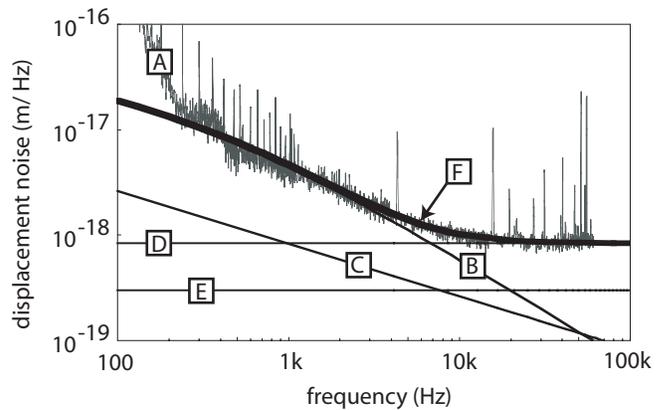}
\caption{\label{fig:result} Displacement noise ($S^{1/2}(\omega)$) in the interferometer with sapphire mirrors as a function of frequency $f=\omega / 2 \pi$. Curve ``A'' is the total measured noise; ``B'' is the calculated thermoelastic-damping noise; ``C'' is the expected coating thermal noise; ``D'' is the measured shot noise; and ``E'' is the electronic noise in the measurement, principally due to the photodetectors. Curve ``F'' is the sum, in quadrature, of the thermoelastic-damping noise and the shot noise. There are no undetermined parameters in these curves. No fits to the data were performed.} 
\end{figure}
\end{center}

Figure~\ref{fig:result} shows a plot of the measured displacement noise $S^{1/2}(\omega)$ in the interferometer, along with predicted curves for the thermoelastic-damping noise and coating thermal noise, and measured values of the shot noise and electronic noise. No parameters were adjusted in the theory to fit the data. We measured the shot noise in our photodetectors by shining a heat lamp on them.

The thermoelastic-damping noise prediction agrees well with the observed noise floor of this instrument over approximately one decade in frequency, from 400 Hz to 5 kHz. The observed noise floor falls slightly below the prediction between about $500 Hz$ and $1 kHz$, and the fit can be improved by adjusting the parameters $\alpha$ and $\kappa$ by approximately $10\%$. This is not surprising, since the values of $\alpha$ and $\kappa$ reported in the literature often vary by this much or more~\cite{Braginsky99-2}. Note that the thermoelastic characteristic frequency, $\omega_c /2 \pi = 134 Hz$, is low enough that we cannot measure the thermoelastic-damping noise below this frequency because of seismic noise. We are, however, still able to observe the transition region between high and low frequency behavior.

It is important to separate coating thermal noise from thermoelastic-damping noise if we are to make a definitive measurement of the latter. We expect the coating noise to fall well below the observed total noise curve based on our previous measurements of the mechanical losses in identical coatings~\cite{Black04-2}. The theoretical model we use to calculate coating noise takes into account the different Young's modulus and Poisson's ratio of the coating and substrate, and it admits the possibility that the coating mechanical loss angle might be different for strains parallel and perpendicular to the substrate-coating interface. For a coating with thickness $d$, Young's modulus $E_c$, and Poisson's ratio $\sigma_c$, the thermal noise is~\cite{Harry02}
\begin{eqnarray}
\delta \ell^2_{coat} (f) &=& \frac{2 k_B T}{\pi^{3/2} f} \frac{1 - \sigma^2}{E w_0} \left\{ \frac{1}{\sqrt{\pi}} \frac{d}{w_0} \frac{1}{E E_c (1-\sigma_c^2)(1-\sigma^2)} \right. \nonumber \\
& \times & \left[ E_c^2 (1+\sigma)^2(1-2 \sigma)^2 \phi_{\parallel} \right. \nonumber \\
&+& E E_c \sigma_c (1+\sigma)(1+\sigma_c)(1-2\sigma)(\phi_{\parallel} - \phi_{\perp}) \nonumber \\
&+& \left. \left. E^2 (1+\sigma_c)^2(1-2 \sigma_c) \phi_{\perp} \right] \right\}
\label{eq:full-coating}
\end{eqnarray}
where $\phi_{\parallel}$ and $\phi_{\perp}$ are the coating's mechanical loss angles for strains parallel and perpendicular to the substrate-coating interface, respectively; $E$ and $\sigma$ are the Young's modulus and Poisson's ratio for the substrate. Table~I gives the values of the parameters used to calculate both the thermoelastic-damping noise and the coating thermal noise for our instrument.

%\newpage

\begin{center}
{\bf Table I: Parameters used to calculate the noise floor of the instrument}
\begin{tabular}{lccl} 
$\alpha$ & $5.1 \times 10^{-6}$ & $K^{-1}$ & \cite{BENCH} \\
$\kappa$ & $36$ &  $W/m \cdot K$ & \cite{BENCH} \\
$\rho$ & $3.983 \times 10^{3}$ &  $kg/m^3$ & \cite{BENCH} \\
$C$ & $770$ & $J/kg K$ & \cite{BENCH} \\
$\sigma$ & $0.23$ && \cite{BENCH} \\
$\sigma_c$ & $0.2$ && \cite{Harry04-T} \\
$w_0$ & $160$ & $\mu m$ & \\
$E$ & $40 \times 10^{10}$ & $N/m^2$ & \cite{Harry02} \\
$E_c$ & $11.0 \times 10^{10}$ &  $N/m^2$ & \cite{Harry02} \\
$d$ & $4.26$ & $\mu m$ & \cite{REO} \\
$\phi_{\parallel}$ & $2.7 \times 10^{-4}$ && \cite{Penn03} \\
$\phi_{\perp}$ & $2.7 \times 10^{-4}$ && \cite{Black04-2} \\
\end{tabular}
\end{center}

%In this table, $\sigma_c$ is the Poisson ratio of the coating, $E$ is the Young's modulus of the substrate, $E_c$ is the Young's modulus of the coating, $d$ is the coating thickness, and $\phi_c$ is the mechanical loss angle of the coating.  We evaluated the coating thermal noise for the sapphire mirrors using the method described in Reference~\cite{Harry02}. We demonstrated in a previous paper that this method is valid for nominally identical coatings on fused-silica substrates~\cite{Black04-2}. 

\section{Summary}

We have constructed a high-sensitivity, Fabry-Perot-based interferometer using suspended optics to measure the intrinsic thermoelastic-damping noise in sapphire mirrors. The theoretical prediction of Cerdonio, et al.~\cite{Cerdonio01} agrees well with our observed noise floor, provided we include the necessary correction to the integral $J ( \Omega )$. This result provides a direct confirmation of the theory used to predict the noise floor of an advanced interferometric gravitational wave detector with sapphire mirrors, thus providing input to the choice of mirror substrates for the next generation of detectors. In addition, this work establishes that, at least at this level of sensitivity (down to $8 \times 10^{-19} m/\sqrt{Hz}$) and in this frequency range (above $150 Hz$), sapphire mirrors exhibit no other, unforseen noise sources.

\section{Acknowledgments} 

This work was supported by the NSF under grant number PHY01-07417.

\bibliographystyle{apsrev}
\bibliography{/Users/eric/Documents/Bibliographies/astrophysics,/Users/eric/Documents/Bibliographies/thermal-noise-story,/Users/eric/Documents/Bibliographies/Photothermal}

\begin{thebibliography}{25}
\expandafter\ifx\csname natexlab\endcsname\relax\def\natexlab#1{#1}\fi
\expandafter\ifx\csname bibnamefont\endcsname\relax
  \def\bibnamefont#1{#1}\fi
\expandafter\ifx\csname bibfnamefont\endcsname\relax
  \def\bibfnamefont#1{#1}\fi
\expandafter\ifx\csname citenamefont\endcsname\relax
  \def\citenamefont#1{#1}\fi
\expandafter\ifx\csname url\endcsname\relax
  \def\url#1{\texttt{#1}}\fi
\expandafter\ifx\csname urlprefix\endcsname\relax\def\urlprefix{URL }\fi
\providecommand{\bibinfo}[2]{#2}
\providecommand{\eprint}[2][]{\url{#2}}

\bibitem[{\citenamefont{Braginsky et~al.}(1999)\citenamefont{Braginsky,
  Gorodetsky, and Vyatchanin}}]{Braginsky99-2}
\bibinfo{author}{\bibfnamefont{V.~B.} \bibnamefont{Braginsky}},
  \bibinfo{author}{\bibfnamefont{M.~L.} \bibnamefont{Gorodetsky}},
  \bibnamefont{and} \bibinfo{author}{\bibfnamefont{S.~P.}
  \bibnamefont{Vyatchanin}}, \bibinfo{journal}{Phys. Lett. A}
  \textbf{\bibinfo{volume}{264}}, \bibinfo{pages}{1} (\bibinfo{year}{1999}).

\bibitem[{\citenamefont{Cerdonio et~al.}(2001)\citenamefont{Cerdonio, Conti,
  Heidmann, and Pinard}}]{Cerdonio01}
\bibinfo{author}{\bibfnamefont{M.}~\bibnamefont{Cerdonio}},
  \bibinfo{author}{\bibfnamefont{L.}~\bibnamefont{Conti}},
  \bibinfo{author}{\bibfnamefont{A.}~\bibnamefont{Heidmann}}, \bibnamefont{and}
  \bibinfo{author}{\bibfnamefont{M.}~\bibnamefont{Pinard}},
  \bibinfo{journal}{Phys. Rev. D} \textbf{\bibinfo{volume}{63}},
  \bibinfo{pages}{082003} (\bibinfo{year}{2001}).

\bibitem[{\citenamefont{Barish and Weiss}(1999)}]{Barish99}
\bibinfo{author}{\bibfnamefont{B.~C.} \bibnamefont{Barish}} \bibnamefont{and}
  \bibinfo{author}{\bibfnamefont{R.}~\bibnamefont{Weiss}},
  \bibinfo{journal}{Physics Today} \textbf{\bibinfo{volume}{52}},
  \bibinfo{pages}{44} (\bibinfo{year}{1999}).

\bibitem[{\citenamefont{L\"{u}ck and {the GEO600 team}}(1997)}]{Luck97}
\bibinfo{author}{\bibfnamefont{H.}~\bibnamefont{L\"{u}ck}} \bibnamefont{and}
  \bibinfo{author}{\bibnamefont{{the GEO600 team}}},
  \bibinfo{journal}{Classical and Quantum Gravity}
  \textbf{\bibinfo{volume}{14}}, \bibinfo{pages}{1471} (\bibinfo{year}{1997}).

\bibitem[{\citenamefont{Caron et~al.}(1997)\citenamefont{Caron, Dominjo,
  Drezen, Flaminio, Grave, and more}}]{Caron97}
\bibinfo{author}{\bibfnamefont{B.}~\bibnamefont{Caron}},
  \bibinfo{author}{\bibfnamefont{A.}~\bibnamefont{Dominjo}},
  \bibinfo{author}{\bibfnamefont{C.}~\bibnamefont{Drezen}},
  \bibinfo{author}{\bibfnamefont{R.}~\bibnamefont{Flaminio}},
  \bibinfo{author}{\bibfnamefont{X.}~\bibnamefont{Grave}}, \bibnamefont{and}
  \bibinfo{author}{\bibfnamefont{M.}~\bibnamefont{more}},
  \bibinfo{journal}{Classical and Quantum Gravity}
  \textbf{\bibinfo{volume}{14}}, \bibinfo{pages}{1461} (\bibinfo{year}{1997}).

\bibitem[{\citenamefont{Kawabe and {the TAMA collaboration}}(1997)}]{Kawabe97}
\bibinfo{author}{\bibfnamefont{K.}~\bibnamefont{Kawabe}} \bibnamefont{and}
  \bibinfo{author}{\bibnamefont{{the TAMA collaboration}}},
  \bibinfo{journal}{Classical and Quantum Gravity}
  \textbf{\bibinfo{volume}{14}}, \bibinfo{pages}{1477} (\bibinfo{year}{1997}).

\bibitem[{\citenamefont{Blair}(2000)}]{Blair00}
\bibinfo{author}{\bibfnamefont{D.~G.} \bibnamefont{Blair}},
  \bibinfo{journal}{General Relativity and Gravitation}
  \textbf{\bibinfo{volume}{32}}, \bibinfo{pages}{371} (\bibinfo{year}{2000}).

\bibitem[{\citenamefont{Buonanno and Chen}(2001)}]{Buonanno01}
\bibinfo{author}{\bibfnamefont{A.}~\bibnamefont{Buonanno}} \bibnamefont{and}
  \bibinfo{author}{\bibfnamefont{Y.}~\bibnamefont{Chen}},
  \bibinfo{journal}{Class. Quantum Grav.} \textbf{\bibinfo{volume}{18}},
  \bibinfo{pages}{L95} (\bibinfo{year}{2001}).

\bibitem[{\citenamefont{Gustafson et~al.}(1999)\citenamefont{Gustafson,
  Shoemaker, Strain, and Weiss}}]{LIGO2}
\bibinfo{author}{\bibfnamefont{E.}~\bibnamefont{Gustafson}},
  \bibinfo{author}{\bibfnamefont{D.}~\bibnamefont{Shoemaker}},
  \bibinfo{author}{\bibfnamefont{K.~A.} \bibnamefont{Strain}},
  \bibnamefont{and} \bibinfo{author}{\bibfnamefont{R.}~\bibnamefont{Weiss}},
  \bibinfo{type}{Internal Working Note} \bibinfo{number}{LIGO-T990080-00-D},
  \bibinfo{institution}{LIGO Project} (\bibinfo{year}{1999}).

\bibitem[{\citenamefont{Rowan et~al.}(2000)\citenamefont{Rowan, Cagnoli,
  Sneddon, Hough, Route, Gustafson, Fejer, and Mitrofanov}}]{Rowan00}
\bibinfo{author}{\bibfnamefont{S.}~\bibnamefont{Rowan}},
  \bibinfo{author}{\bibfnamefont{G.}~\bibnamefont{Cagnoli}},
  \bibinfo{author}{\bibfnamefont{P.}~\bibnamefont{Sneddon}},
  \bibinfo{author}{\bibfnamefont{J.}~\bibnamefont{Hough}},
  \bibinfo{author}{\bibfnamefont{R.}~\bibnamefont{Route}},
  \bibinfo{author}{\bibfnamefont{E.~K.} \bibnamefont{Gustafson}},
  \bibinfo{author}{\bibfnamefont{M.~M.} \bibnamefont{Fejer}}, \bibnamefont{and}
  \bibinfo{author}{\bibfnamefont{V.}~\bibnamefont{Mitrofanov}},
  \bibinfo{journal}{Phys. Lett. A} \textbf{\bibinfo{volume}{265}},
  \bibinfo{pages}{5} (\bibinfo{year}{2000}).

\bibitem[{\citenamefont{Ju et~al.}(1996)\citenamefont{Ju, Notcutt, Blair,
  Bondu, and Zhao}}]{Ju96}
\bibinfo{author}{\bibfnamefont{L.}~\bibnamefont{Ju}},
  \bibinfo{author}{\bibfnamefont{M.}~\bibnamefont{Notcutt}},
  \bibinfo{author}{\bibfnamefont{D.}~\bibnamefont{Blair}},
  \bibinfo{author}{\bibfnamefont{F.}~\bibnamefont{Bondu}}, \bibnamefont{and}
  \bibinfo{author}{\bibfnamefont{C.~N.} \bibnamefont{Zhao}},
  \bibinfo{journal}{Phys. Lett. A} \textbf{\bibinfo{volume}{218}},
  \bibinfo{pages}{197} (\bibinfo{year}{1996}).

\bibitem[{\citenamefont{Saulson}(1994)}]{Saulson-book}
\bibinfo{author}{\bibfnamefont{P.~R.} \bibnamefont{Saulson}},
  \emph{\bibinfo{title}{Fundamentals of Interferometric Gravitational Wave
  Detectors}} (\bibinfo{publisher}{World Scientific},
  \bibinfo{address}{Singapore}, \bibinfo{year}{1994}).

\bibitem[{\citenamefont{Onsager}(1931{\natexlab{a}})}]{Onsager31-1}
\bibinfo{author}{\bibfnamefont{L.}~\bibnamefont{Onsager}},
  \bibinfo{journal}{Phys. Rev.} \textbf{\bibinfo{volume}{37}},
  \bibinfo{pages}{405} (\bibinfo{year}{1931}{\natexlab{a}}).

\bibitem[{\citenamefont{Onsager}(1931{\natexlab{b}})}]{Onsager31-2}
\bibinfo{author}{\bibfnamefont{L.}~\bibnamefont{Onsager}},
  \bibinfo{journal}{Phys. Rev.} \textbf{\bibinfo{volume}{38}},
  \bibinfo{pages}{2265} (\bibinfo{year}{1931}{\natexlab{b}}).

\bibitem[{\citenamefont{Callen and Welton}(1951)}]{Callen51}
\bibinfo{author}{\bibfnamefont{H.~B.} \bibnamefont{Callen}} \bibnamefont{and}
  \bibinfo{author}{\bibfnamefont{T.~A.} \bibnamefont{Welton}},
  \bibinfo{journal}{Phys. Rev.} \textbf{\bibinfo{volume}{83}},
  \bibinfo{pages}{34} (\bibinfo{year}{1951}).

\bibitem[{\citenamefont{Callen and Greene}(1952)}]{Callen52}
\bibinfo{author}{\bibfnamefont{H.~B.} \bibnamefont{Callen}} \bibnamefont{and}
  \bibinfo{author}{\bibfnamefont{R.~F.} \bibnamefont{Greene}},
  \bibinfo{journal}{Phys. Rev.} \textbf{\bibinfo{volume}{86}},
  \bibinfo{pages}{702} (\bibinfo{year}{1952}).

\bibitem[{\citenamefont{Ch\^{a}telier}(1884)}]{LeChatelier84}
\bibinfo{author}{\bibfnamefont{H.~L.~L.} \bibnamefont{Ch\^{a}telier}},
  \bibinfo{journal}{Comptes rendus.} \textbf{\bibinfo{volume}{99}},
  \bibinfo{pages}{786} (\bibinfo{year}{1884}).

\bibitem[{\citenamefont{Saulson}(1990)}]{Saulson90}
\bibinfo{author}{\bibfnamefont{P.~R.} \bibnamefont{Saulson}},
  \bibinfo{journal}{Phys. Rev. D} \textbf{\bibinfo{volume}{42}},
  \bibinfo{pages}{2437} (\bibinfo{year}{1990}).

\bibitem[{\citenamefont{Numata}({31 Dec. 2003})}]{Numata03}
\bibinfo{author}{\bibfnamefont{K.}~\bibnamefont{Numata}},
  \bibinfo{journal}{Phys. Rev. Lett.} \textbf{\bibinfo{volume}{91}},
  \bibinfo{pages}{260602} (\bibinfo{year}{{31 Dec. 2003}}).

\bibitem[{\citenamefont{Black et~al.}(2004)\citenamefont{Black, Villar,
  Barbary, Bushmaker, Heefner, Kawamura, Kawazoe, Matone, Meidt, Rao
  et~al.}}]{Black04-2}
\bibinfo{author}{\bibfnamefont{E.~D.} \bibnamefont{Black}},
  \bibinfo{author}{\bibfnamefont{A.}~\bibnamefont{Villar}},
  \bibinfo{author}{\bibfnamefont{K.}~\bibnamefont{Barbary}},
  \bibinfo{author}{\bibfnamefont{A.}~\bibnamefont{Bushmaker}},
  \bibinfo{author}{\bibfnamefont{J.}~\bibnamefont{Heefner}},
  \bibinfo{author}{\bibfnamefont{S.}~\bibnamefont{Kawamura}},
  \bibinfo{author}{\bibfnamefont{F.}~\bibnamefont{Kawazoe}},
  \bibinfo{author}{\bibfnamefont{L.}~\bibnamefont{Matone}},
  \bibinfo{author}{\bibfnamefont{S.}~\bibnamefont{Meidt}},
  \bibinfo{author}{\bibfnamefont{S.~R.} \bibnamefont{Rao}},
  \bibnamefont{et~al.}, \bibinfo{journal}{Phys. Lett. A}
  \textbf{\bibinfo{volume}{328}}, \bibinfo{pages}{1} (\bibinfo{year}{2004}).

\bibitem[{REO()}]{REO}
\emph{\bibinfo{title}{Research electro-optics}},
  \bibinfo{note}{http://www.reoinc.com}.

\bibitem[{\citenamefont{Harry et~al.}(2002)\citenamefont{Harry, Gretarsson,
  Saulson, Kittelberger, Penn, Startin, Rowan, Fejer, Crooks, Cagnoli
  et~al.}}]{Harry02}
\bibinfo{author}{\bibfnamefont{G.~M.} \bibnamefont{Harry}},
  \bibinfo{author}{\bibfnamefont{A.~M.} \bibnamefont{Gretarsson}},
  \bibinfo{author}{\bibfnamefont{P.~R.} \bibnamefont{Saulson}},
  \bibinfo{author}{\bibfnamefont{S.~E.} \bibnamefont{Kittelberger}},
  \bibinfo{author}{\bibfnamefont{S.~D.} \bibnamefont{Penn}},
  \bibinfo{author}{\bibfnamefont{W.~J.} \bibnamefont{Startin}},
  \bibinfo{author}{\bibfnamefont{S.}~\bibnamefont{Rowan}},
  \bibinfo{author}{\bibfnamefont{M.~M.} \bibnamefont{Fejer}},
  \bibinfo{author}{\bibfnamefont{D.~R.~M.} \bibnamefont{Crooks}},
  \bibinfo{author}{\bibfnamefont{G.}~\bibnamefont{Cagnoli}},
  \bibnamefont{et~al.}, \bibinfo{journal}{Class. Quantum Grav.}
  \textbf{\bibinfo{volume}{19}}, \bibinfo{pages}{897} (\bibinfo{year}{2002}).

\bibitem[{\citenamefont{Finn}()}]{BENCH}
\bibinfo{author}{\bibfnamefont{L.~S.} \bibnamefont{Finn}},
  \emph{\bibinfo{title}{{BENCH}: {S}cience-grounded figures of merit for
  comparing interferometric gravitational wave detector designs}},
  \bibinfo{note}{{O}pen-source interferometer characterization program, v. 2.0;
  {http://ligo.mit.edu/$\sim$pf/Bench/}.}

\bibitem[{\citenamefont{Harry}(2004)}]{Harry04-T}
\bibinfo{author}{\bibfnamefont{G.}~\bibnamefont{Harry}}, \bibinfo{type}{Tech.
  Rep.} \bibinfo{number}{{T040029-00-R}}, \bibinfo{institution}{LIGO}
  (\bibinfo{year}{2004}).

\bibitem[{\citenamefont{Penn et~al.}(2003)\citenamefont{Penn, Sneddon,
  Armandula, Betzweiser, Cagnoli, Camp, Crooks, Fejer, M., Harry
  et~al.}}]{Penn03}
\bibinfo{author}{\bibfnamefont{S.~D.} \bibnamefont{Penn}},
  \bibinfo{author}{\bibfnamefont{P.~H.} \bibnamefont{Sneddon}},
  \bibinfo{author}{\bibfnamefont{H.}~\bibnamefont{Armandula}},
  \bibinfo{author}{\bibfnamefont{J.~C.} \bibnamefont{Betzweiser}},
  \bibinfo{author}{\bibfnamefont{G.}~\bibnamefont{Cagnoli}},
  \bibinfo{author}{\bibfnamefont{J.}~\bibnamefont{Camp}},
  \bibinfo{author}{\bibfnamefont{D.~R.~M.} \bibnamefont{Crooks}},
  \bibinfo{author}{\bibfnamefont{M.~M.} \bibnamefont{Fejer}},
  \bibinfo{author}{\bibfnamefont{G.~A.} \bibnamefont{M.}},
  \bibinfo{author}{\bibfnamefont{G.~M.} \bibnamefont{Harry}},
  \bibnamefont{et~al.}, \bibinfo{journal}{Class. Quantum Grav.}
  \textbf{\bibinfo{volume}{20}}, \bibinfo{pages}{2917} (\bibinfo{year}{2003}).

\end{thebibliography}

\end{document}